# Thickness-shear Frequencies of an Infinite Quartz Plate with Material Property Variation Along the Thickness

Ji Wang[*], Wenliang Zhang, Dejin Huang, Tingfeng Ma, Jianke Du
Piezoelectric Device Laboratory, School of Mechanical Engineering & Mechanics, Ningbo University,
818 Fenghua Road, Ningbo, Zhejiang 315211, CHINA
[*]E-mail: wangji@nbu.edu.cn

*Abstract*—Properties of the quartz crystal blank of a resonator is assumed homogeneous, uniform, and perfect in design, manufacturing, and applications. As end products, quartz crystal resonators are frequently exposed to gases and liquids which can cause surface damage and internal degradation of blanks under increasingly hostile conditions. The combination of service conditions and manufacturing process including chemical etching and polishing can inevitably modify the surface of quartz crystal blanks with changes of material properties, raising the question of what will happen to vibrations of quartz crystal resonators of thickness-shear type if such modifications to blanks are to be evaluated for sensitive applications. Such questions have been encountered in other materials and structures with property variations either on purpose or as the effect of environmental or natural processes commonly referred to as functionally graded materials, or FGMs. Analyses have been done in applications as part of studies on FGMs in structural as well as in acoustic wave device applications. A procedure based on series solutions has been developed in the evaluation of frequency changes and features in an infinite quartz crystal plate of AT-cut with the symmetric material variation pattern given in a cosine function with the findings that the vibration modes are now closely coupled. These results can be used in the evaluation of surface damage and corrosion of quartz crystal blanks of resonators in sensor applications or development of new structures of resonators.

## I. Introduction

In the analysis and design of quartz crystal resonators, it has been known that the material is homogeneous and uniform naturally. This provides the single analytical procedure required in the vibration analysis of anisotropic plates with high frequency modes and particularly strong couplings of the thickness and flexural ones. The difficulties in the accurate analysis of anisotropic plate vibrations are generally known with complications from anisotropic material, coupled modes, and accompanying boundary conditions. These complications have been there and many techniques primarily based on approximation have been developed for solutions which are required in resonator design. Further refinement and improvement of analysis are made with the consideration of more physical complications like irregular configurations of crystal blanks such as the commonly known beveling, which refers to the contours or thickness variations in edges of blanks to suppress strong couplings between thickness-shear and flexural modes. It has been found that the beveling in common configurations such as linear or quadratic variation of thickness can weaken couplings between thickness-shear and flexural modes significantly [1-2]. As a result, various processing techniques have been developed to take the advantage by making contours or beveling in edges of crystal blank in most products.

In vibration analysis with the Mindlin plate equations, thickness variation of a plate is equivalent to changes of elastic constants and stiffness. This means we may be able to realize the reduction of mode couplings through changes of elastic constants in certain regions of the crystal plate blank, as the precise beveling intended to achieve. Clearly, further studies on the precise pattern and scheme of variation of material properties, which is also called the functionally graded materials (FGMs), in the quartz crystal resonators, are needed in finding alternative method in quartz crystal resonator processing and manufacturing. In addition to take the advantage of FGM in resonator design and fabrication, there is also a practical concern that the quartz crystal resonators exposed in gas and liquid will result in surface and internal damages which will also cause material properties variation with possibly equal effects of FGMs. The changes in functions and properties of resonators may not be neglected due to its sensitive nature or for the evaluation of service conditions. Obviously, it is equally important to study the pattern of material property change and effects on the performance and properties of resonators. Furthermore, such studies may provide valid guidelines of FGMs in quartz crystal resonators and proper pattern of FGM can be produced to improve the performance of new generations of resonators.

There have been studies on wave propagations in FGM solids with various methods including approximations and discrete techniques [3-4]. There is no doubt that such analysis is important in establishing the correlations between performance changes and FGM patterns in wave propagation and vibrations of device structures. In the material processing part, FGM patterns can be achieved through modern technologies such as laser radiation, chemical etching, and so on. We can make the partial materials to FGM materials so advantages can be taken and the equivalent effect of beveling can be achieved with minimal cost on processing. In this case, we have finally utilized the FGMs for possible advantages rather than known applications in structural protections and enhancements [5]. Of course, applications of FGMs in acoustic wave devices have been studied before for possible performance enhancements and novel fabrication techniques [6-9]. The essential nature of acoustic wave devices requires the analysis to consider wave propagations or high frequency vibrations in piezoelectric materials and solids as have been done for FGM plates and structures [6-14]. Such studies have been carried out before and there are positive leads to be followed up with more research for effective FGMs to meet performance requirements of next generation of acoustic wave devices.

## II. PROBLEM AND FORMULATION

We start with an infinite plate of quartz crystal with modified material properties resembling to FGMs. The plate is shown in Fig.1, and the thickness of plate is $2b$.

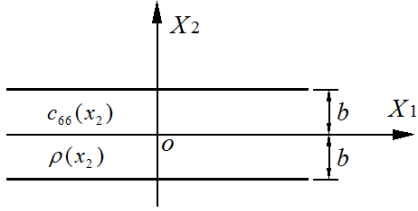

Fig. 1  An infinite quartz crystal plate

For the simplest bulk acoustic waves (BAW) propagating in the elastic plate, the stress equation of motion is

$$(c_{66}u_{,2})_{,2} = \rho(x_2)\ddot{u}, \quad (1)$$

where $c_{66}$, $\rho$, and $u$ are elastic constant, density, and displacements of plate, respectively.

In case of $c_{66}$ and $\rho$ are constants, the simple solution is known as

$$u = A\sin\frac{x_2}{2b}e^{i\omega t}, \quad (2)$$

$$\omega_n = \frac{n\pi}{2b}\sqrt{\frac{c_{66}}{\rho}}, n = 1,2,\cdots, \quad (3)$$

where $A$ and $\omega$ are amplitudes and vibration frequency, respectively.

In this study, we assume that quartz crystal plate has been treated and constants $c_{66}$ and $\rho$ are functions of the thickness coordinate $x_2$ in the form of

$$c_{66} = C_{66}\cos\frac{x_2}{Nb}, \quad (4)$$

$$\rho = \rho_0\cos\frac{x_2}{Nb}, \quad (5)$$

where $C_{66}$ and $\rho_0$ are constants, and $N$ is assumed to be a large integer to ensure the smooth variation along the thickness.

Apparently, if $N$ is really large, then $c_{66}$ and $\rho$ are close constant. We want to examine the difference of vibrations caused by the small variation of $c_{66}$ and $\rho$ through appropriate choice of $N$.

With the consideration of traction-free boundary conditions of the plate in Fig. 1, we can assume the thickness displacement as

$$u = \sum_{n=1,3,5,\cdots}^{\infty} a_n \sin\frac{n\pi}{2b}x_2 e^{i\omega t}. \quad (6)$$

Then through the expansion of (1) we have

$$c_{66,2}u_{,2} + c_{66}u_{,22} = \rho(x_2)\ddot{u}, \quad (7)$$

and by substituting (6) into (7) we obtain

$$-\frac{1}{Nb}C_{66}\sin\frac{x_2}{Nb}\sum_{n=1,3,5,\cdots}^{\infty} a_n \frac{n\pi}{2b}\cos\frac{n\pi}{2b}x_2$$

$$-C_{66}\cos\frac{x_2}{Nb}\sum_{n=1,3,5,\cdots}^{\infty} a_n \left(\frac{n\pi}{2b}\right)^2 \sin\frac{n\pi}{2b}x_2$$

$$= -\rho_0\cos\frac{x_2}{Nb}\sum_{n=1,3,5,\cdots}^{\infty} \omega^2 a_n \sin\frac{n\pi}{2b}x_2. \quad (8)$$

To simplify (8), we can use

$$\sin\frac{x_2}{Nb}\cos\frac{n\pi}{2b}x_2 = \cos\frac{x_2}{Nb}\sum_{m=1,3,5,\cdots}^{\infty} b_{nm}\sin\frac{m\pi}{2b}x_2. \quad (9)$$

This actually is to obtain the Fourier expansion of function $\tan\frac{x_2}{Nb}\cos\frac{n\pi}{2b}x_2$ in sine series in the form of

$$\tan\frac{x_2}{Nb}\cos\frac{n\pi}{2b}x_2 = \sum_{m=1,3,5,\cdots}^{\infty} b_{nm}\sin\frac{m\pi}{2b}x_2, \quad (10)$$

where $b_{nm}$ is the Fourier coefficient in the form of

$$b_{nm} = \frac{1}{b}\int_{-b}^{b}\tan\frac{x_2}{Nb}\cos\frac{n\pi}{2b}x_2\sin\frac{m\pi}{2b}x_2\,dx_2$$

$$= \int_{0}^{1}\tan\frac{X}{N}\left[\sin\frac{m+n}{2}\pi X + \sin\frac{m-n}{2}\pi X\right]dX. \quad (11)$$

By substituting (9) into the first term of (7), we have

$$-\frac{1}{Nb}C_{66}\sin\frac{x_2}{Nb}\sum_{n=1,3,5,\cdots}^{\infty} a_n\frac{n\pi}{2b}\cos\frac{n\pi}{2b}x_2$$

$$= -\frac{1}{Nb}C_{66}\cos\frac{x_2}{Nb}\sum_{n=1,3,5,\cdots}^{\infty} a_n\frac{n\pi}{2b}\sum_{m=1,3,5,\cdots}^{\infty} b_{nm}\sin\frac{m\pi}{2b}x_2$$

$$= -\frac{1}{Nb}\frac{\pi}{2b}C_{66}\cos\frac{x_2}{Nb}\sum_{m=1,3,5,\cdots}^{\infty}(a_1 b_{1m} + 3a_3 b_{3m} + \cdots$$

$$+ na_n b_{nm} + \cdots)\sin\frac{m\pi}{2b}x_2. \quad (12)$$

Then substituting (12) into (7), we have

$$\frac{1}{Nb}\frac{\pi}{2b}C_{66}\cos\frac{x_2}{Nb}\sum_{m=1,3,5,\cdots}^{\infty}(a_1 b_{1m} + 3a_3 b_{3m} + \cdots + na_n b_{nm}$$

$$+\cdots)\sin\frac{m\pi}{2b}x_2 + C_{66}\cos\frac{x_2}{Nb}\sum_{n=1,3,5,\cdots}^{\infty}a_n\left(\frac{n\pi}{2b}\right)^2\sin\frac{n\pi}{2b}x_2$$

$$= \rho_0\cos\frac{x_2}{Nb}\sum_{n=1,3,5,\cdots}^{\infty}\omega^2 a_n\sin\frac{n\pi}{2b}x_2. \quad (13)$$

Finally (13) can be simplified as

$$\left(\frac{\pi}{2Nb^2}C_{66}b_{11} + \frac{\pi^2}{4b^2}C_{66} - \rho_0\omega^2\right)a_1 + \frac{3\pi}{2Nb^2}C_{66}b_{31}a_3 +$$

$$\cdots + \frac{\pi}{2Nb^2}C_{66}nb_{n1}a_n = 0,$$

$$\frac{\pi}{2Nb^2}C_{66}b_{13}a_1 + \left(\frac{3\pi}{2Nb^2}C_{66}b_{33} + \frac{9\pi^2}{4b^2}C_{66} - \rho_0\omega^2\right)a_3 +$$

$$\cdots + \frac{\pi}{2Nb^2}C_{66}nb_{n3}a_n = 0,$$

$$\vdots$$

$$\frac{\pi}{2Nb^2}C_{66}b_{1n}a_1 + \frac{3\pi}{2Nb^2}C_{66}b_{3n}a_3 + \cdots + \left(\frac{\pi}{2Nb^2}C_{66}nb_{nn}\right.$$

$$\left. + \frac{n^2\pi^2}{4b^2}C_{66} - \rho_0\omega^2\right)a_n = 0. \quad (14)$$

With the normalized parameters

$$\Omega = \frac{\omega}{\omega_1}, \Omega^2 = \frac{\omega^2}{\omega_1^2} = \rho_0\omega^2\frac{4b^2}{\pi^2 C_{66}}, \quad (15)$$

where $\omega_1$ is the fundamental thickness-shear frequency of a plate with uniform properties given in (3), then we can rewrite (14) as

$$\left(\frac{2b_{11}}{N\pi} + 1^2 - \Omega^2\right)a_1 + \frac{6b_{31}}{N\pi}a_3 + \cdots + \frac{2nb_{n1}}{N\pi}a_n = 0,$$

$$\frac{2b_{13}}{N\pi}a_1 + \left(\frac{6b_{33}}{N\pi} + 3^2 - \Omega^2\right)a_3 + \cdots + \frac{2nb_{n3}}{N\pi}a_n = 0,$$

$$\vdots$$

$$\frac{2b_{1n}}{N\pi}a_1 + \frac{6b_{3n}}{N\pi}a_3 + \cdots + \left(\frac{2nb_{nn}}{N\pi} + n^2 - \Omega^2\right)a_n = 0. \quad (16)$$

Now we have a system of equations for amplitudes of displacements as

$$[K]*\{A\} = 0, \quad (17)$$

where $[K]$ and $\{A\}$ are matrices as

$$\{A\} = \{a_1 \quad a_3 \quad \cdots \quad a_n\}^T, \quad (18)$$

$$[K] = \begin{bmatrix} \frac{2b_{11}}{N\pi}+1-\Omega^2 & \frac{6b_{31}}{N\pi} & \cdots & \frac{2nb_{n1}}{N\pi} \\ \frac{2b_{13}}{N\pi} & \frac{6b_{33}}{N\pi}+3^2-\Omega^2 & \cdots & \frac{2nb_{n3}}{N\pi} \\ \vdots & \vdots & \ddots & \vdots \\ \frac{2b_{1n}}{N\pi} & \frac{6b_{3n}}{N\pi} & \cdots & \frac{2nb_{nn}}{N\pi}+n^2-\Omega^2 \end{bmatrix}. \quad (19)$$

For free vibrations of an infinite FGM quartz crystal plate, vibration frequencies can be obtained by setting the determinant of $K$ matrix to vanish through

$$|K| = 0. \quad (20)$$

With all known parameters of material properties, we can evaluate (20) for the fundamental and overtone vibration frequencies. It is clear from (19) that the fundamental and overtone modes are now closely coupled. In other words, the FGM plate can no longer support pure modes of thickness-shear vibrations even it is infinite. In this case, our interests will be on the effect of FGM patterns on frequencies of the coupled thickness-shear modes.

### III. NUMERICAL RESULTS

As mentioned before, vibration frequencies $\Omega_n(n = 1,2,\cdots,\infty)$ can be calculated from (16). In case only two vibration modes need to be considered, or $n$ is set to 2, we have the analytical solution as

$$\Omega = \sqrt{5 + \frac{b_{11}+3b_{33}}{N\pi} \mp \sqrt{16 - \frac{8(b_{11}-3b_{33})}{N\pi} + \frac{(b_{11}-3b_{33})^2}{N^2\pi^2} + \frac{12b_{13}b_{31}}{N^2\pi^2}}}. \quad (21)$$

If $n$ is greater than 2, frequency solutions cannot be given explicitly. In any case, we have to make a decision about the cut-off frequency in the calculation and subsequently decide how many terms are to be included in the determinant. In the beginning, probably we can only carry out calculations based on trials using the accuracy of frequency solutions as criteria for the termination of calculation. The increase of numbers of terms in the matrix will improve the accuracy of frequency solutions until their convergence. For this reason, we need to decide the frequency range, or the cut-off frequency, from the beginning of the calculation.

The number $N$ in above equations will also have significant effect on the determination of frequency solutions. For really large $N$, the FGM pattern will be almost uniform, thus we can expect solutions with weakly coupled overtone modes. For smaller $N$, the FGM pattern will be significant and we expect to see great changes on overtone mode frequencies. For $N$ in between, there will strong couplings of modes and clear effects on the frequency solutions. As an example, frequency solutions for different FGM patterns and cut-off frequencies are given in Tables I-IX below based on convergent results of different $N$ and number of terms. It is important to point out that we have had a series of tests to examine the relationship between cut-off frequency and number of terms to be included in the calculation. Such results can be used for better estimation about number of terms to be included in future calculations.

TABLE I.  VIBRATION FREQUENCIES WITH THE ORDER OF DETERMINANT = 1

| Mode no. | $N=5$ | $N=10$ | $N=15$ | $N=20$ | $N=25$ |
|---|---|---|---|---|---|
| 1 | 1.004066 | 1.001014 | 1.000450 | 1.000253 | 1.000162 |

TABLE II.  VIBRATION FREQUENCIES WITH THE ORDER OF DETERMINANT = 2

| Mode no. | $N=5$ | $N=10$ | $N=15$ | $N=20$ | $N=25$ |
|---|---|---|---|---|---|
| 1 | 1.004075 | 1.001015 | 1.000451 | 1.000253 | 1.000162 |
| 3 | 3.001365 | 3.000339 | 3.000150 | 3.000084 | 3.000054 |

TABLE III.  VIBRATION FREQUENCIES WITH THE ORDER OF DETERMINANT = 3

| Mode no. | $N=5$ | $N=10$ | $N=15$ | $N=20$ | $N=25$ |
|---|---|---|---|---|---|
| 1 | 1.004076 | 1.001015 | 1.000451 | 1.000253 | 1.000162 |
| 3 | 3.001374 | 3.000339 | 3.000150 | 3.000085 | 3.000054 |
| 5 | 5.000815 | 5.000203 | 5.000090 | 5.000051 | 5.000032 |

TABLE IV.  VIBRATION FREQUENCIES WITH THE ORDER OF DETERMINANT = 4

| Mode no. | $N=5$ | $N=10$ | $N=15$ | $N=20$ | $N=25$ |
|---|---|---|---|---|---|
| 1 | 1.004076 | 1.001015 | 1.000451 | 1.000253 | 1.000162 |
| 3 | 3.001376 | 3.000339 | 3.000150 | 3.000085 | 3.000054 |
| 5 | 5.000825 | 5.000204 | 5.000090 | 5.000051 | 5.000032 |
| 7 | 7.000579 | 7.000145 | 7.000064 | 7.000036 | 7.000023 |

TABLE V.  VIBRATION FREQUENCIES WITH THE ORDER OF DETERMINANT = 5

| Mode no. | $N=5$ | $N=10$ | $N=15$ | $N=20$ | $N=25$ |
|---|---|---|---|---|---|
| 1 | 1.004076 | 1.001015 | 1.000451 | 1.000253 | 1.000162 |
| 3 | 3.001376 | 3.000339 | 3.000150 | 3.000085 | 3.000054 |
| 5 | 5.000826 | 5.000204 | 5.000090 | 5.000051 | 5.000032 |
| 7 | 7.000589 | 7.000145 | 7.000064 | 7.000036 | 7.000023 |
| 9 | 9.000448 | 9.000112 | 9.000050 | 9.000028 | 9.000018 |

TABLE VI.  VIBRATION FREQUENCIES WITH THE ORDER OF DETERMINANT = 6

| Mode no. | $N=5$ | $N=10$ | $N=15$ | $N=20$ | $N=25$ |
|---|---|---|---|---|---|
| 1 | 1.004077 | 1.001015 | 1.000451 | 1.000253 | 1.000162 |
| 3 | 3.001376 | 3.000339 | 3.000150 | 3.000085 | 3.000054 |
| 5 | 5.000826 | 5.000204 | 5.000090 | 5.000051 | 5.000032 |
| 7 | 7.000590 | 7.000145 | 7.000064 | 7.000036 | 7.000023 |
| 9 | 9.000457 | 9.000113 | 9.000050 | 9.000028 | 9.000018 |

TABLE VII.  VIBRATION FREQUENCIES WITH THE ORDER OF DETERMINANT = 7

| Mode no. | $N=5$ | $N=10$ | $N=15$ | $N=20$ | $N=25$ |
|---|---|---|---|---|---|
| 1 | 1.004077 | 1.001015 | 1.000451 | 1.000253 | 1.000162 |
| 3 | 3.001376 | 3.000339 | 3.000150 | 3.000085 | 3.000054 |
| 5 | 5.000826 | 5.000204 | 5.000090 | 5.000051 | 5.000032 |
| 7 | 7.000590 | 7.000145 | 7.000064 | 7.000036 | 7.000023 |
| 9 | 9.000459 | 9.000113 | 9.000050 | 9.000028 | 9.000018 |

TABLE VIII.  VIBRATION FREQUENCIES WITH THE ORDER OF DETERMINANT = 8

| Mode no. | $N=5$ | $N=10$ | $N=15$ | $N=20$ | $N=25$ |
|---|---|---|---|---|---|
| 1 | 1.004077 | 1.001015 | 1.000451 | 1.000253 | 1.000162 |
| 3 | 3.001376 | 3.000339 | 3.000150 | 3.000085 | 3.000054 |
| 5 | 5.000826 | 5.000204 | 5.000090 | 5.000051 | 5.000032 |
| 7 | 7.000590 | 7.000145 | 7.000064 | 7.000036 | 7.000023 |
| 9 | 9.000459 | 9.000113 | 9.000050 | 9.000028 | 9.000018 |

TABLE IX.  VIBRATION FREQUENCIES WITH THE ORDER OF DETERMINANT = 9

| Mode no. | $N=5$ | $N=10$ | $N=15$ | $N=20$ | $N=25$ |
|---|---|---|---|---|---|
| 1 | 1.004077 | 1.001015 | 1.000451 | 1.000253 | 1.000162 |
| 3 | 3.001376 | 3.000339 | 3.000150 | 3.000085 | 3.000054 |
| 5 | 5.000826 | 5.000204 | 5.000090 | 5.000051 | 5.000032 |
| 7 | 7.000590 | 7.000145 | 7.000064 | 7.000036 | 7.000023 |
| 9 | 9.000459 | 9.000113 | 9.000050 | 9.000028 | 9.000018 |

By examining the tables above, we found that for specified $N$ of 5, 10, 15, 20, and 25, we can obtain accurate solutions of the first five frequencies with number of terms larger than 7. We used this rule of convergence estimation for the calculation of frequencies given in Table X below.  One clear observation is that the overtone frequencies are no longer the exact multiples of the fundamental frequency.  As the FGM approaches to homogeneous plates, we see the known pattern of frequency change again.

TABLE X.  EFFECTS OF FGM GRADING ON PLATE VIBRATION FREQUENCIES

| FGM index | 1 | 3 | 5 | 7 | 9 |
|---|---|---|---|---|---|
| $N=5$ | 1.004077 | 3.001376 | 5.000827 | 7.000590 | 9.000459 |
| $N=10$ | 1.001015 | 3.000339 | 5.000204 | 7.000145 | 9.000113 |
| $N=15$ | 1.000451 | 3.000150 | 5.000090 | 7.000064 | 9.000050 |
| $N=20$ | 1.000253 | 3.000085 | 5.000051 | 7.000036 | 9.000028 |
| $N=25$ | 1.000162 | 3.000054 | 5.000032 | 7.000023 | 9.000018 |
| $N=\infty$ | 1.000000 | 3.000000 | 5.000000 | 7.000000 | 9.000000 |

Proceedings of 2014 IEEE International Frequency Control Symposium, May 19-22, Taipei International Convention Center, Taipei, Taiwan## IV. CONCLUSIONS

With an infinite FGM quartz crystal plate, we analyzed the thickness-shear vibrations with couplings of the fundamental and thickness-shear overtone modes. This is a unique feature of FGM plates because the modes are not coupled in homogeneous plates. The solutions are obtained by decoupling the equations through Fourier expansion of some coefficients of the differential equation. To represent a plate with FGM material in simple and proper function, we choose the change of density and elastic constants as a cosine function, which preserves the essential feature of symmetry and ensure smooth changes along the thickness. This is a reasonable choice for a more general description of FGM patterns of quartz crystal plates under various treating and damage situations. We can expect that the laser radiation, corrosion by liquids, and long term damage through harmful environments can be well represented with this pattern. The frequency solutions, or frequency deviation from the fundamental and overtone frequencies of homogenous plates then can be used for the evaluation of the damage to plates. The multiple solutions of frequencies then can be used for a coupled estimation which will be more accurate. The results show that the vibration modes of the fundamental and overtone, both symmetric and anti-symmetric, are no longer separated and the couplings are in place for all modes. The neat nature of overtone frequencies also disappeared, and the frequencies are no longer the exact multiples of the fundamental frequency. In addition to evaluate the surface damage and corrosion of material property for resonator and sensor applications, we can also use the results for possible design of FGM quartz crystal resonators with superior properties.

## ACKNOWLEDGMENT

This research is supported by the National Natural Science Foundation of China through grants 10932004 and 11372145.## REFERENCES

[1] P. C. Y. Lee and Ji Wang, "Piezoelectrically forced thickness‐shear and flexural vibrations of contoured quartz resonators," Journal of Applied Physics, vol. 79, no. 7, pp. 3411-3422, 1996.

[2] Ji Wang, P. C. Y. Lee, and D. H. Bailey, "Thickness-shear and flexural vibrations of linearly contoured crystal strips with multiprecision computation," Computers & Structures, vol. 70, no. 4, pp. 437-445, 1999.

[3] Liming Gao, Ji Wang, Zheng Zhong and Jianke Du, "An analysis of surface acoustic wave propagation in functionally graded plates with homotopy analysis method," Acta Mechanica, vol. 208, pp. 249-258, 2009.

[4] Liming Gao, Ji Wang, Zheng Zhong and Jianke Du, "An exact analysis of surface acoustic waves in a plate of functionally graded materials," IEEE Trans. Ultrason., Ferroelect, and Freq. Contr., vol. 52, no. 12, pp. 2693-2700, 2009.

[5] Niino Masayuki, Hirai Toshio, and Watanabe Ryuzo, "Functionally gradient materials. In pursuit of super heat resisting materials for spacecraft," (in Japanese), Journal of the Japan Society for Composite Materials, vol. 13, no. 6, pp. 257-264, 1987.

[6] Wilfredo Montealegre Rubio, Em íio Carlos Nelli Silva, and Julio Cezar Adamowski, "Design of resonators based on functionally graded piezoelectric materials," XII International Symposium on Dynamic Problems of Mechanics, Ilhabela, Brazil, Feb. 26 – Mar. 2, 2007.

[7] Yangyang Chen, Ji Wang, Jianke Du, and Jiashi Yang, "Characterization of material property gradient in a functionally graded material using an AT-cut quartz thickness-shear mode resonator," Philosophical Magazine Letters, vol. 93, no. 6, pp. 362-370, 2013.

[8] Zheng Zhong and Tao Yu, "Vibration of a simply supported functionally graded piezoelectric rectangular plate," Smart Mater. Struct., vol. 15, pp. 1404–1412, 2006.

[9] Trung-Kien Nguyen, Karam Sab, and Guy Bonnet, "First-order shear deformation plate models for functionally graded materials," Composite Structures, vol. 83, pp. 25-36, 2008.

[10] Hiroyuki Matsunaga, "Free vibration and stability of functionally graded plates according to a 2-D higher-order deformation theory", Composite Structures, vol. 82, pp. 499-512, 2008.

[11] Guojun Nie and Zheng Zhong, "Dynamic analysis of multi-directional functionally graded annular plates," Applied Mathematical Modelling, vol. 34, pp. 608–616, 2010.

[12] Sh. Hosseini-Hashemi, M. Fadaee, and S.R. Atashipour, "A new exact analytical approach for free vibration of Reissner–Mindlin functionally graded rectangular plates," International Journal of Mechanical Sciences, vol. 53, pp. 11–22, 2011.

[13] S. Natarajan, P.M. Baiz, S. Bordas, T. Rabczuk, and P. Kerfriden, "Natural frequencies of cracked functionally graded material plates by the extended finite element method," Composite Structures, vol. 93, pp. 3082–3092, 2011.

[14] A. M. A. Neves, A. J. M. Ferreira, E. Carrera, M. Cinefra, C. M. C. Roque, R. M. N. Jorge, and C. M. M. Soares, "Static, free vibration and buckling analysis of isotropic and sandwich functionally graded plates using a quasi-3D higher-order shear deformation theory and a meshless technique," Composites: Part B, vol. 44, no. 1, pp. 657–674, 2013.